\documentclass[10pt,conference]{IEEEtran}
\IEEEoverridecommandlockouts
\usepackage{cite}
\usepackage{amsmath,amssymb,amsfonts}
\usepackage{algorithm, algorithmic}
\usepackage{graphicx}
\usepackage{tabularx,booktabs}
\usepackage{textcomp}
\usepackage{xcolor}
\usepackage{url}
\usepackage{multirow}
\usepackage{enumitem}
\usepackage{color, colortbl}
\usepackage[percent]{overpic}

\definecolor{codegreen}{rgb}{0,0.6,0}
\definecolor{codegray}{rgb}{0.5,0.5,0.5}
\definecolor{codepurple}{rgb}{0.58,0,0.82}
\definecolor{backcolour}{rgb}{0.95,0.95,0.92}
\definecolor{codekeyword}{RGB}{153, 51, 153}

\usepackage{listings}
\lstdefinestyle{mystyle}{
  backgroundcolor=\color{white},   commentstyle=\bfseries\color{codegreen},
  keywordstyle=\bfseries\color{codekeyword},
  numberstyle=\tiny\color{codegray},
  stringstyle=\color{codepurple},
  basicstyle=\ttfamily\scriptsize\linespread{1.15},
  breakatwhitespace=false,         
  breaklines=true,                 
  captionpos=b,                    
  keepspaces=true,                 
  numbers=left,                    
  numbersep=5pt,                  
  showspaces=false,                
  showstringspaces=false,
  showtabs=false,                  
  tabsize=2,
  xleftmargin=16pt,
  xrightmargin=0pt,
  breakindent=0pt,
  resetmargins=true
}
\usepackage[framemethod=TikZ]{mdframed}
\mdfdefinestyle{style1}{
innerleftmargin=0.4cm,innerrightmargin=0.4cm,
innertopmargin=0.3cm ,innerbottommargin=0.3cm,
roundcorner=10pt,linewidth=1.0pt,
footnoteinside=false}

\def\BibTeX{{\rm B\kern-.05em{\sc i\kern-.025em b}\kern-.08em
    T\kern-.1667em\lower.7ex\hbox{E}\kern-.125emX}}
    
\newcolumntype{Y}{>{\centering\arraybackslash}X}

\begin{document}

\title{
Boundary Value Exploration for Software Analysis}

\author{\IEEEauthorblockN{Felix Dobslaw, Francisco Gomes de Oliveira Neto, Robert Feldt}
\IEEEauthorblockA{\textit{Chalmers, and the University of Gothenburg} \\
\textit{Dept.\ of Computer Science and Engineering}\\
Gothenburg, Sweden \\
dobslaw@chalmers.se, francisco.gomes@cse.gu.se, robert.feldt@chalmers.se}
}

\maketitle
\thispagestyle{plain}
\pagestyle{plain}

\begin{abstract}
For software to be reliable and resilient, it is widely accepted that tests must be created and maintained alongside the software itself. One safeguard from vulnerabilities and failures in code is to ensure correct behavior on the boundaries between the input space sub-domains. So-called boundary value analysis (BVA) and boundary value testing (BVT) techniques aim to exercise those boundaries and increase test effectiveness. However, the concepts of BVA and BVT themselves are not generally well defined, and it is not clear how to identify relevant sub-domains, and thus the boundaries delineating them, given a specification. This has limited adoption and hindered automation. We clarify BVA and BVT and introduce Boundary Value Exploration (BVE) to describe techniques that support them by helping to detect and identify boundary inputs. Additionally, we propose two concrete BVE techniques based on information-theoretic distance functions: (i) an algorithm for boundary detection and (ii) the usage of software visualization to explore the behavior of the software under test and identify its boundary behavior. As an initial evaluation, we apply these techniques on a much used and well-tested date handling library. Our results reveal questionable behavior at boundaries highlighted by our techniques. In conclusion, we argue that the boundary value exploration that our techniques enable is a step towards automated boundary value analysis and testing, fostering their wider use and improving test effectiveness and efficiency. 
\end{abstract}


\section{Introduction}
Even though boundary value analysis\slash testing~\cite{white1980domain,clarke1982close,ostrand1988category,reid1997empirical} is a core technique in software testing, it has been acknowledged in the literature that establishing and maintaining correct\slash meaningful behavior at boundaries requires creativity and is hard to realize\cite{bath2012software}. In practice, many faults can be found near boundaries that delimit different sets of inputs that should or are handled differently by the software, i.e., where the behavior of the software changes or should change considerably.

The traditional way to proceed is for testers to start from a specification and then use partition analysis (PA) to partition the input space into partitions or sub-domains\footnote{The latter term is preferred since this allows for sub-domains to (partly) overlap. In contrast, the term `partitions' typically implies they cannot.} in which the behavior of the software should be the same (so-called equivalence partitions) or similar. Then, boundary value analysis (BVA) techniques instruct testers to sample from the boundaries between the sub-domains to obtain and execute tests (BVT) that can ensure correct behavior at the boundaries~\cite{reid1997empirical}. 


A downside of the techniques used for BVA\slash BVT is that they do not give concrete methods for identifying sub-domains; often, it is assumed that the specification explicitly states them and directly helps identify boundaries. However, this might depend on how the specification is written, and since most non-trivial software specifications are not complete, are implicit, or are specified only in natural language and thus vague, it is not clear how to proceed. In particular, this is a problem for complex software with large input spaces, for heavily or non-linearly dependent inputs, and when inputs are of complex and highly structured types. Overall, the dated establishment of BVA\slash BVT and their scope limits their more extensive use for software testing and quality assurance. Furthermore, since this process typically depends on human analysis, it is not clear, in general, how to automate it, which limits efficiency.

In this paper, we \textit{introduce the concept of boundary value exploration (BVE)} to support BVA and BVT in situations where the specification might \textit{not} be complete, consistent, explicit, or even exist. We propose two concrete techniques for BVE based on information-theoretic distance functions previously offered for measuring test diversity~\cite{feldt2008searching,feldt2016test,feldt2019towards}. For illustration and evaluation, we then apply these techniques on a much used and well-tested component for the handling of dates in the high-level programming language Julia\footnote{\url{https://julialang.org/}}. Our results reveal questionable behavior at some of the identified boundaries. 
In particular, our boundary detection algorithm showed the SUT's boundaries by measuring and visualizing the values of the program derivative~\cite{feldt2019towards}. In short, our contributions are that we:

\begin{itemize}
    \item propose a clarification of BVA and BVT and their relation, and
    \item introduce the novel concept of Boundary Value Exploration (BVE) to support BVA and BVT by identifying candidate boundary values, and
    \item describe two concrete techniques for BVE, respectively, a boundary detection algorithm using distance functions as a means for detection, and software visualization to enhance the understanding of boundaries and select additional boundary inputs.
\end{itemize}

The rest of this paper is structured as follows. Section \ref{rw} presents an overview of the most relevant existing literature on BVA and BVT. In Section \ref{bva}, we clarify and distinguish BVA and BVT, introduce BVE and describe their relation, followed by a motivating example highlighting the prospects using BVE in Section~\ref{example}. In Section \ref{method} we then further describe and apply the BVE techniques for boundary detection and visualization to our case and reveal boundary candidates.
Section \ref{discussion} briefly discusses the findings and their implications for software testing while Section \ref{conclusions} concludes.

\section{Background and Related Work}
\label{rw}

In equivalence (or category\slash domain) partitioning and testing, similar test inputs are grouped in equivalence classes covering contiguous regions of the input space\slash domain~\cite{white1980domain,clarke1982close,ostrand1988category,reid1997empirical}. The assumption is that similar inputs would be handled in a similar way\footnote{In the original formulations, similarity referred solely to the program taking the same execution path for all inputs in the same partition\slash domain. Still, over time the terms have been used in a more general way.}, e.g., the program would go through the same path by the software under test and be more likely to be affected by the same faults. Thus, testing one of the inputs in a partition should be enough to catch such faults~\cite{ostrand1988category,chen2010art}. 

There are many partition strategies, but sampling test inputs for each partition are typically still challenging and costly. Strategies based on search algorithms~\cite{marculescu2018finding} or combinatorial testing~\cite{moran2018mapreduce} can help. Still, there is a fundamental trade-off between the cost of sampling tests\slash inputs and the cost of running and evaluating the results~\cite{arcuri2012random}. 
For instance, Adaptive Random Testing (ART) techniques use distance measures to sample test inputs further apart from each other~\cite{chen2010art} to cover different partitions and increase test effectiveness. However, the high cost of repeatedly calculating distances often hinder applicability in realistic systems, such that simple random exploration of the input space can be more effective~\cite{arcuri2012random}.  
Even though random testing techniques can generate many test inputs, the effective coverage of partitions remains prohibitive if the execution of sampled tests is slow or an automated oracle is unavailable~\cite{arcuri2012random}. 
An alternative approach would be to sample at or around the boundaries between partitions since it is critical to ensure that inputs on either side of a boundary are correctly handled, so-called boundary value analysis (BVA) or testing (BVT)~\cite{reid1997empirical,hierons2006avoiding}.

Traditionally, BVA (or BVT the terms have often been used interchangeably) is based on a human developer analyzing a specification to identify partitions and the boundary values they imply and then write down test cases that ensure correct behavior at the boundaries~\cite{reid1997empirical}. While early results questioned the value of partitioning and BVA in comparison to random testing~\cite{hamlet1990partition}, later results showed that BVA could have higher fault-detecting ability than both random testing and equivalence partitioning (EP)~\cite{reid1997empirical}. However, neither EP nor BVA has been defined in a generalizable way, since they rely on a non-formal understanding of what constitutes a partition. Alternative ways of coming up with boundary values have been used, e.g., based on experience from previously problematic inputs, formerly known issues, or ``natural'' boundaries of the data types involved~\cite{marculescu2018finding}, analysis of source code~\cite{zhang2015bvacode}, or based on distance\slash diversity metrics~\cite{kim2019guiding,marculescu2018finding}. 
However, regardless of the method and agent used (e.g., a human tester or automation via an algorithm such as search) to identify boundary values, few techniques actively seek and, once found, explore the boundary to exploit it for fault-revealing tests~\cite{marculescu2018finding,kim2019guiding}. One exception is Kim et al.~\cite{kim2019guiding}, who propose using distance values between activation traces of deep neural networks (DNN) to guide the identification of test inputs.
These values are used to identify the test input able to ``surprise'' the DNN, hence, e.g., image classification, revealing which images could exercise the classification boundaries. Similar ideas were discussed previously in~\cite{poulding2017generating}, where distance functions were used to identify boundaries between valid and invalid inputs of a SUT and then in~\cite{marculescu2018finding} to mutate them to follow such boundaries systematically. This allowed the automated creation of test cases for robustness testing. However, the approach focuses on the boundaries between valid and invalid inputs, not considering more general boundaries within the space of valid inputs.

Hierons in \cite{hierons2006avoiding} describes the challenges of using BVA in systems that produce the expected outputs but for the wrong reasons (due to accidental correctness). Consequently, BVA was adapted to select test inputs with specific properties, such as choosing them from partitions that yield different outputs or utilizing metrics based on the observed test behavior. More generally, Feldt and Dobslaw in ~\cite{feldt2019towards} propose a metric based on information theory that combines input and output distances to detect areas of maximum ``change'', i.e., a derivative in mathematical parlance. They even propose to use the derivative as a formal definition of a boundary for BVA. Their proposal suggests that boundaries can be detected without the need for an oracle. Instead, universal information-theoretic metrics previously proposed for test diversity and based on compression can be utilized as general distance functions~\cite{feldt2008searching,feldt2016test} if no specification- or data type-specific distance functions are known~\cite{feldt2019towards}.
However, the example application of~\cite{feldt2019towards} only uses exhaustive search, which is not feasible for testing realistic software with, often, substantial input spaces.



The approaches mentioned above do not use the identified boundaries as sources of information to enhance BVA and BVT, such as seeking different neighborhoods or areas of invalid values in the input space. It is challenging to handle non-contiguous valid spaces~\cite{marculescu2018finding} or understand the constraints behind multiple boundaries~\cite{zhang2015bvacode}. Our approach complements the benefits of BVA by fostering and emphasizing the exploration of such boundaries. It also proposes concrete methods to detect the boundaries in the first place, allowing automation and automated BVT test creation. In particular, our methods augment the developer's\slash tester's capabilities seeking and using the boundary values based on software visualization.

In general, software visualization (SV) tools and techniques focus primarily on the visualization of structure, behavior, and dependencies of software artifacts, processes among other elements of software development~\cite{Bedu2019_vissoft,Mattila2016_slr}.
Literature reports on several tools and techniques in SV that enhances software comprehensibility when applied to architecture~\cite{Shanin2014_architecture}, maintenance~\cite{Koschke2003_slr}, evolution~\cite{Salameh2016_slr}, and even when applied to more general domains that do not necessarily target a specific area of software engineering~\cite{Bedu2019_vissoft}.

In connection to testing, most SV proposed tools and techniques focus on maintenance, fault detection, or change impact analysis~\cite{Mattila2016_slr}. Authors in~\cite{deOliveiraNeto2018visualizing} propose similarity maps as a visualization technique that uses dimensionality reduction to visualize the diversity between tests in a test suite. 
In their industrial evaluation, the similarity maps brought awareness to relevant issues related to maintenance of the tests where unnecessary redundancy was being added and kept to test repositories introducing waste~\cite{deOliveiraNeto2018visualizing}.
Similarly, authors in~\cite{deOliveiraNeto2018_issre} developed a tool that mines test execution logs in search of details about failing tests (e.g., error messages, exceptions thrown, timestamps). This information is then aggregated into distinct classes of failures and then displayed in dashboards in Continuous Integration monitoring systems. Their evaluation with industry partners reveals that their dashboards provide a holistic view of the failures and aids testers identify the faults behind the failures~\cite{deOliveiraNeto2018_issre}.

Feldt et al.~\cite{feldt2013supporting} apply software visualization to test history data to support more effective analysis, planning, and quality assurance execution. Mainly, test information is visualized in a heatmap and monitored through meetings with stakeholders. Their results show that practitioners can use the heatmaps to prioritize test effort, resource allocation, or awareness of the SUT's problematic areas. Moreover, the authors analyze the correlation between the heatmap and the different development data (e.g., code churn or the number of failures) to highlight stakeholders' attention areas. Nonetheless, these and other visualization tools must collaborate with stakeholders to support data-driven decision making in software development~\cite{feldt2013supporting}. Engstr\"om et al.~ \cite{engstrom2014supporting} investigate further the usage of heatmaps from test history to support decision making. Their evaluation reveals that practitioners find the visualization useful to support test planning; however, the type of visualization required is dependent on the task, and participants reported on the importance of interaction with it.

Furthermore, Borg et al.~\cite{borg2018analytical} combine the test results from a project with the design's file structure under test into an interactive 3D city visualization (i.e., a cityscape). Their tool shows a landscape of the various files committed into a project as a building, where the building's height is the number of times the file was committed, and a color gradient indicates how often the committed file, respectively, failed or passed. The resulting visualization reveals insights about regression testing activities such as error-prone areas and tests that should be changed to increase coverage. 

As these SV studies show, visualization can help developers and tester get an overview, trigger reflection, and spot essential patterns in the testing and quality of the software being studied. Similarly, we also use software visualization to trigger insights and reflection on boundary behavior by showing candidate boundary values. This can then help test generation, planning, and, generally, decision making. In particular, our use of visualization with interaction to explore boundary areas, via 3D plots of the input space, is novel. 




\section{Boundary Value Analysis, Testing, and Exploration}
\label{bva}
Boundary value analysis (BVA) and boundary value testing (BVT) are umbrella terms to describe different techniques that identify and ensure correct software behavior at boundaries. While the former is typically presented as a black-box technique focused on identifying partitions and, thus, boundaries were given a specification, the latter is seen as a white-box technique to ensure that the boundary is where it is supposed to be. However, the two terms are often used interchangeably without clarifying how they differ or overlap~\cite{feldt2019towards}.
In the following, we define these concepts more clearly, relate them to each other, and propose a new concept, Boundary Value Exploration (BVE), as a set of techniques that can support them both.



A useful characterization was given by Hierons in~\cite{hierons2006avoiding}, which defined boundaries more formally through their elements, namely pairs of input $(x_1,x_2)$ for adjacent sub-domains $S_1$ and $S_2$ with $x_1 \in S_1$, $x_2 \in S_2$, and $x_1$ and $x_2$ being close together. The latter criterion requires that an ordering or ideally a metric has been identified~\cite{hierons2006avoiding}. Recently, Feldt and Dobslaw noted~\cite{feldt2019towards} that generally applicable compression distances from information theory can be used in this context and by also considering the distance between the outputs, i.e. $d_{output}(o_1, o_2)$ where $o_1 = P(x_1)$ and $o_2 = P(x_2)$ and where $P(x_i)$ denotes running the program $P$ on the input $x_i$, they proposed that boundaries can be defined as pairs that lead to high values of the program derivative $d_{output}(o_1, o_2) / d_{input}(x_1, x_2)$\footnote{Here, $d_{output}$ and $d_{input}$ are two distance functions, one for outputs and one for inputs. While they can be the same, and compressions-based distances like the normalized compression distance can be good default choices~\cite{feldt2008searching,feldt2016test}, they need not be, and a tester can select specific and multiple distance functions depending on their needs and their knowledge of the specification and or the implementation.}. 

A problem with previous proposals for BVA has been that they assume sub-domains to exist, but they do not describe how to identify sub-domains given a specification. Even when a complete, formal specification is available, it might not be clear how to identify its sub-domains. The problem is further exacerbated since, in practice, specifications are often implicit (undocumented) or, even when explicit (documented), they are frequently incomplete, hard to understand, or even incorrect. The proposal of program derivative~\cite{feldt2019towards} allows a clear alternative: define boundaries to be subsets of pairs of inputs with high derivative values for relevant distance functions on inputs and outputs. By selecting relevant input pairs and running an implementation on them and then calculating the distance and derivative values, we can explore actual boundaries in an implementation.

\begin{figure}
    \centering
    \includegraphics[width=\columnwidth]{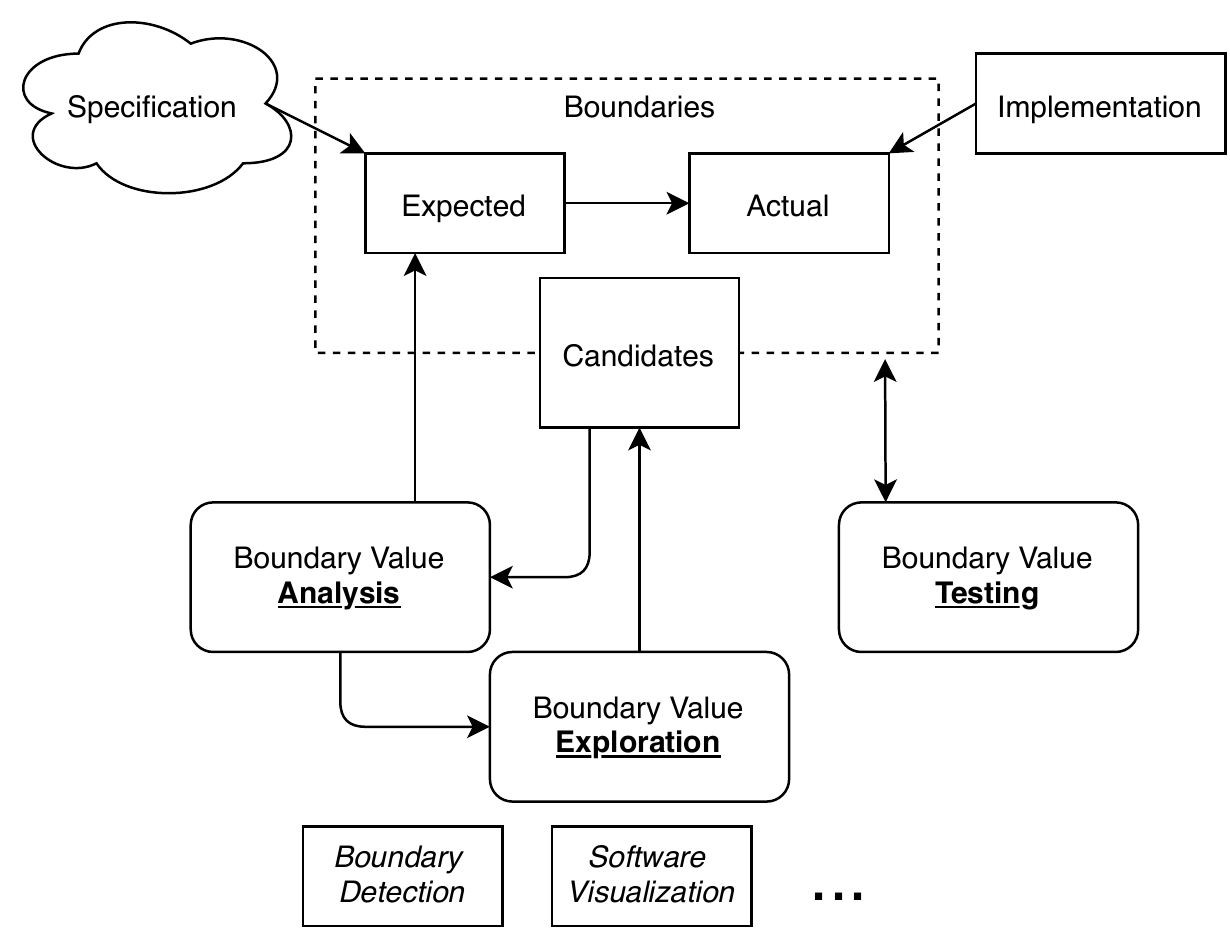}
    \caption{Overview of our approach to Boundary Value Analysis and Testing supported by Boundary Value Exploration techniques. This shows how the key activities (rounded corners) on the bottom relate to the primary artifacts (sharp corners) on top and the central relations between them.}
    \label{fig:bve_approach}
\end{figure}

Figure~\ref{fig:bve_approach} shows an overview of our approach by outlining the primary artifacts (on top, regular rectangles), their activities (lower half, rounded rectangles), and their direct relations (arrows). 
Central to our approach are the boundaries of which we highlight two main types: expected boundaries that are either clear from a specification or which a boundary value analysis identifies, and actual boundaries arising from the current implementation. Boundary value testing is then defined as the design and execution of test cases that compare whether the expected and actual boundaries coincide. While this requires that both expected and actual boundaries are known, it can also help refine the knowledge about both types of boundaries (thus, the arrow is not uni-directional).

Our approach, in Figure~\ref{fig:bve_approach}, also acknowledges that understanding about boundaries can come from multiple sources and not only from analysis of a specification. If the execution of a test during BVT shows a discrepancy between an expected and actual boundary, this might not always indicate a problem in the implementation; we might need to refine our understanding of, and thus update the description of, the expected boundaries. However, \textit{this paper's main novelty is the proposal that techniques for Boundary Value Exploration (BVE) can help refine the knowledge of boundaries by proposing candidate boundary values}. These candidate boundary values can prompt BVA and help identify additional or refine existing expected boundaries. While we define 

\begin{itemize}
\item BVA as the analysis of artifacts of the software development process to clarify the expected and actual boundaries of a piece of software, and 
\item BVT as the execution of specific input pairs in order to ensure that an actual boundary is also expected, we define 
\item BVE as a collection of techniques that select or help select inputs to detect and identify boundary candidates.
\end{itemize}

While we envision that many different techniques for exploring boundaries are applicable, we will focus on techniques based on the recently proposed program derivative~\cite{feldt2019towards} since it was specifically designed to quantify areas of abrupt changes in software behavior. It can be viewed as a concrete definition of boundaries and is a prime candidate for their automatic detection and exploration.

\section{Motivating Example}
\label{example}
To demonstrate the mechanics and impact of BVE in practice and motivate its use, we here present and apply two concrete BVE techniques, namely boundary detection, and boundary value visualization, on a concrete case. We have chosen the \texttt{Date} library of the Julia programming language as our investigation object since it is relatively straightforward while still showing interesting behavior. 

Julia offers a useful feature for the in-code retrieval of type minimum and maximum values for numeric types and composite types based on numeric types using the \texttt{typemin} and \texttt{typemax} functions. \texttt{Date} is one such type as it is typically constructed from three integers, the year, month, and day. For instance, in Julia version 1.1.1, we can apply \texttt{typemax} to the \texttt{Date} type:

\vspace{-.2cm}

{\footnotesize{
\begin{align*}
&\texttt{typemax}(\texttt{Date}) \rightarrow 252522163911149\text{-}12\text{-}31
\end{align*}}}

\vspace{-.5cm}

This result's validity can be verified through the instantiation of a date from its three integer parameters. Table \ref{tab:date} contains various such date instantiations identified during our boundary value exploration and which we will refer to in the following. The above example can be found in row 1, for example.

Since \texttt{typemax(Date)} should, by construction, be an extreme value of type \texttt{Date}, we would expect later dates to be invalid. However, when testing, it turns out the next date is valid and seemingly correct, as shown in row 2 of Table \ref{tab:date}. However, if we continue and increment the year parameter beyond the typemax value while keeping month and day unchanged, this results in a seemingly invalid, but non-exceptional, result (see row 3). The function \texttt{Dates.day} for that date returns 28, whereas 31 was entered\footnote{Worse still, for the month field, an invalid value is assigned.}. Our exploration has helped demonstrating that \texttt{Date} in Julia 1.1.1 is not isomorph in its API, since 

\begin{equation*}
\footnotesize
d == \texttt{Date}(\texttt{Dates.year}(d), \texttt{Dates.month}(d),\texttt{Dates.day}(d))
\end{equation*}

does not hold for all its instances. Still, inconsistent with the above, overstepping the month or day parameters through the \texttt{Date} instantiating function does trigger exceptions, as shown in rows 10 and 11 of Table \ref{tab:date}.

\begin{table*}
    \centering
    \scriptsize
    \caption{Boundary candidates for various input values identified during our boundary value exploration of the Julia 1.1.1 Date API.}
    \label{tab:date}
    \begin{tabularx}{\textwidth}{llll}
    \toprule
        \textbf{Nr.} & \textbf{Command}  & \textbf{Output} & \textbf{Explanation} \\
        \midrule
        1&\texttt{Date(}252522163911149,12,31) & \texttt{252522163911149-12-31} & \texttt{typemax(Date)}\\
        2&\texttt{Date(}252522163911150,1,1) & \texttt{252522163911150-01-01} & \texttt{typemax(Date)} + 1 day\\
        3&\texttt{Date(}252522163911150,12,31) & \texttt{-252522163911150-6028347736506387-28} & \texttt{typemax(Date)} + 1 year\\
        4&\texttt{Date(}252522163911151,10,7) & \texttt{252522163911151-10-07} & \texttt{typemax(Date)} + 280 days\\
        5&\texttt{Date(}252522163911151,10,8) & \texttt{-252522163911150-6028347736506385-06} & \texttt{typemax(Date)} + 281 days\\
        6&\texttt{Date(}-252522163911150,1,1) & \texttt{-252522163911150-01-01} &
        \texttt{typemin(Date)}\\
        7&\texttt{Date(}-252522163911151,12,31) & \texttt{-252522163911151-12-31} & \texttt{typemin(Date)} - 1 day\\
        8&\texttt{Date(}-252522163911151,7,25) & \texttt{-252522163911151-07-25} & \texttt{typemin(Date)} - 161 days\\
        9&\texttt{Date(}-252522163911151,7,24) & \texttt{252522163911150--6028347736506379--07} & \texttt{typemin(Date)} - 162 days\\
        10&\texttt{Date(}2020,12,\textbf{32}) & \texttt{\footnotesize ERROR: ArgumentError: Day: 32 out of range (1:31)...} & day out of range\\
        11&\texttt{Date(}2020,\textbf{0},31) & \texttt{\footnotesize ERROR: ArgumentError: Month: 0 out of range (1:12)...} & month out of range\\
        12&\texttt{Date(typemax(Int)},1,1) &
        \texttt{63131837319416-12056695473012772--7378697629483820630} & year is \texttt{typemax(Int)}\\
        \bottomrule
    \end{tabularx}
\end{table*}

Such acceptance of overflow values can cause runtime errors in the software. Even proper error handling in the calling code does not help due to accepting an invalid state. In practice, possible triggers for resulting runtime errors are invalid user input, e.g., by adversaries, or the reading of erroneous files.

Consequentially, the \texttt{Date} implementation seems to deviate from the \texttt{typemax} specification. In the Julia type \texttt{Date} example, and others, there may be practical reasons for the deviation of boundaries, e.g., those of performance. Either way, the boundaries of \texttt{typemin} and \texttt{typemax} are ill-defined, to say the least, and cannot be relied upon in code.

The question is what the \textit{actual} boundaries of type \texttt{Date} in Julia are and how they can be detected? It turns out that with the help of diversity measures in support of information theory, we can identify the actual boundaries through the in this paper proposed use of BVE without the need for an oracle, domain knowledge, or access to the code. This is demonstrated below.

\begin{figure}
    \centering
    \includegraphics[width=\columnwidth]{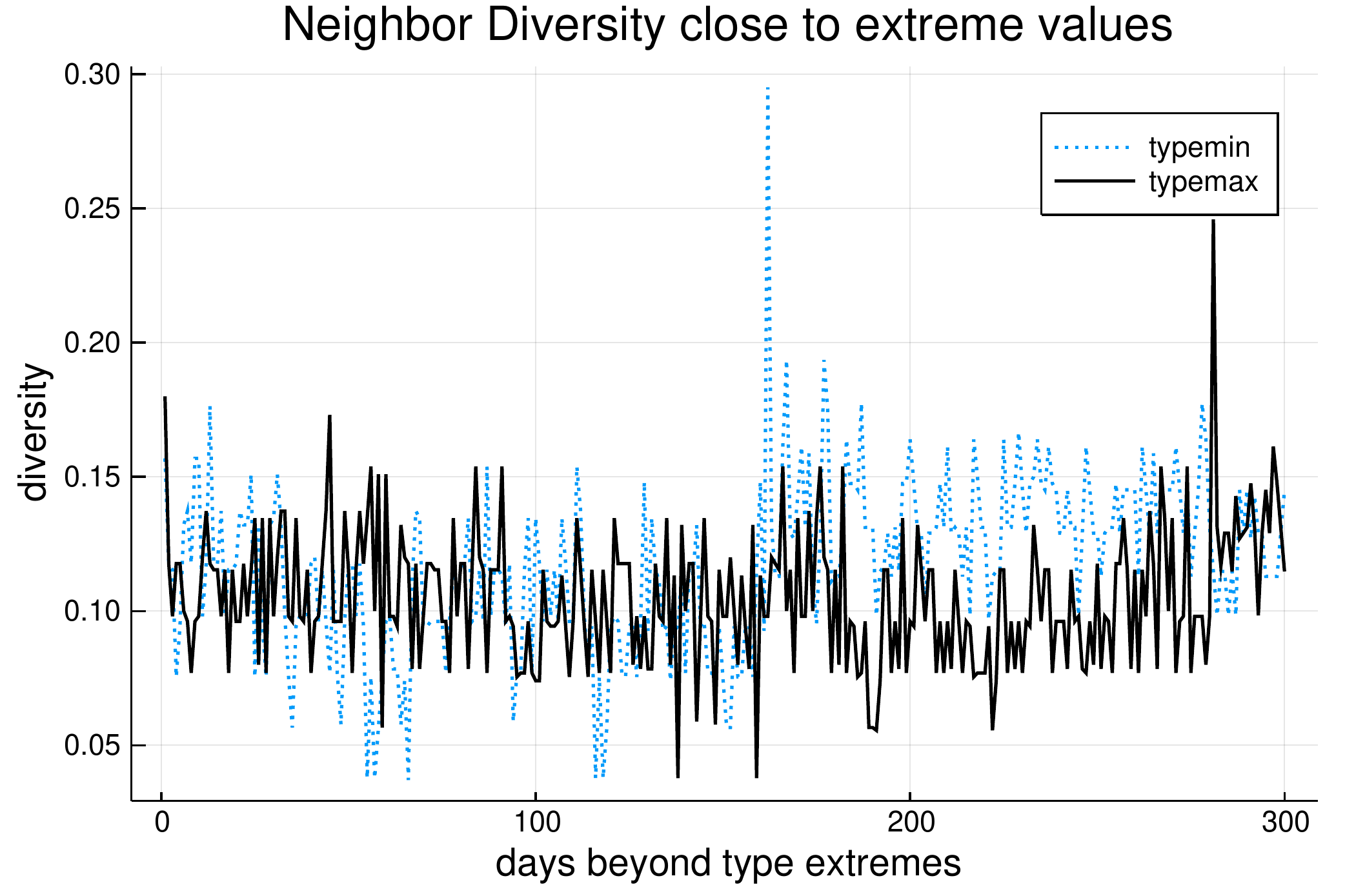}
    \caption{The calculated diversity of consecutive days for Julia \texttt{Date} instantiation starting at \texttt{typemin} (dotted line) and \texttt{typemax} (solid line). Both progress beyond the specified boundaries, meaning for \texttt{typemin} back in time, and for \texttt{typemax} forward. A broad diversity is expected at the on-set for a specified extreme value, but not observed for either case. However, extreme values that are potential boundary candidates and of interest for further investigation (for \texttt{typemin} at 162 and 281 for \texttt{typemax}) exist. The BD-algorithm introduced in \ref{method} extracts these outliers for that purpose.}
    \label{fig:extremes}
\end{figure}

\begin{figure}
    \centering
    \includegraphics[width=\columnwidth]{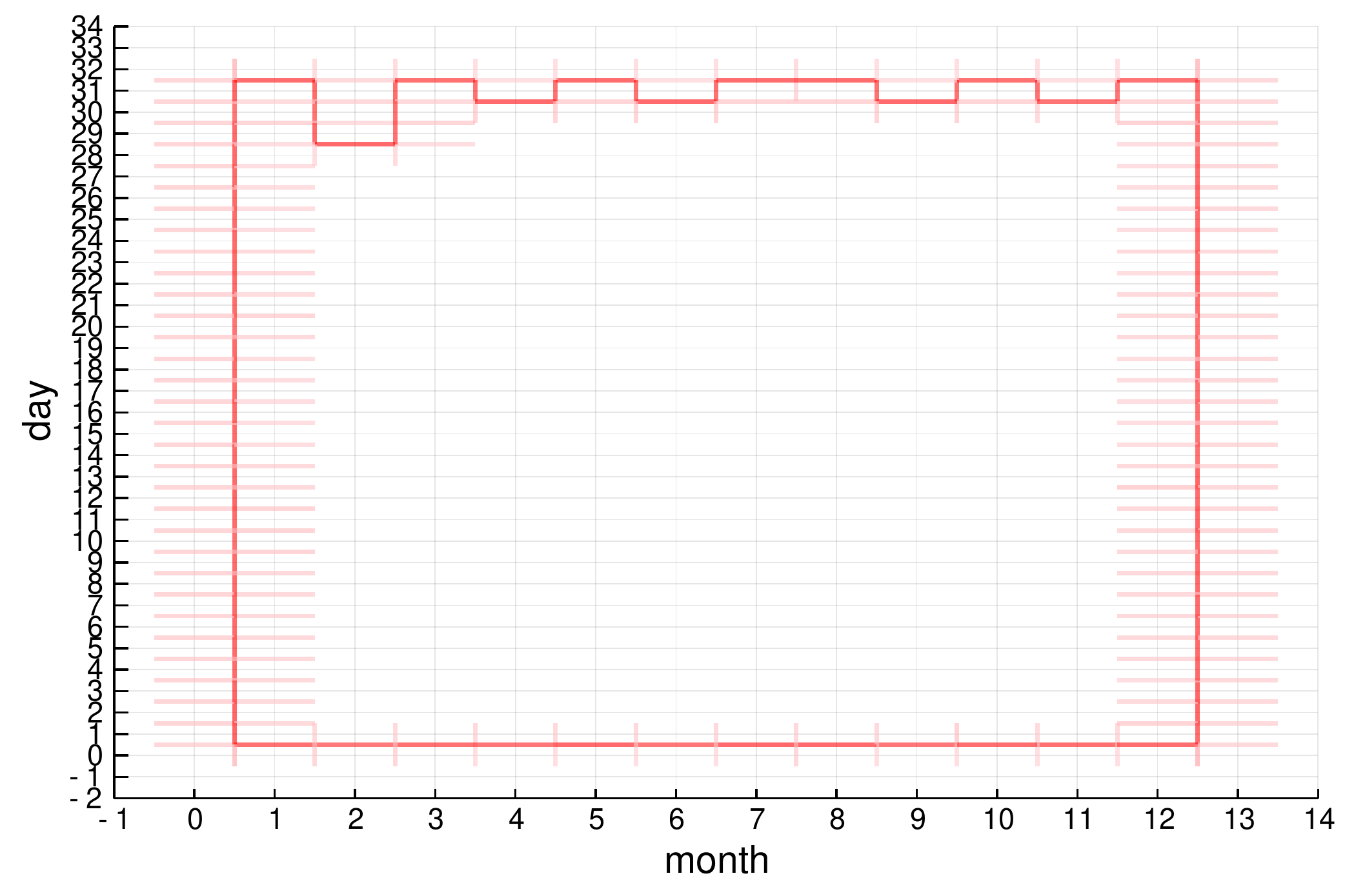}
    \caption{Without an oracle, we can extract a boundary of the software. Here we see a boundary of the Date API for a non-leap year, with input dimensions \textit{day} and \textit{month}.}
    \label{fig:date2d}
\end{figure}

Nearby input/output pairs that deviate strongly suggest boundaries. We would expect pairs nearby the specified \texttt{typemin} and \texttt{typemax} boundaries to peak in neighbor diversity relative to other valid/non-extreme neighborhood pairs. Figures \ref{fig:extremes} and \ref{fig:date2d} exemplify how diversity information may be used to learn about the actual boundaries, again on Julia \texttt{Date}. Figure \ref{fig:extremes} illustrates the diversity for neighboring pairs starting from the extremes \texttt{typemin} (backward in time) and \texttt{typemax} (forward in time). What can be observed is that not aligned with expectations, the derivatives around both extremes are not outstanding. In contrast, outlier peaks in diversity can be observed further out in the supposedly invalid space for both cases. Intuitively, these peaks represent uncommonly diverse neighbors, which suggests boundaries for further exploration. We call those outliers here boundary candidates.

\begin{figure}
\centering
 \begin{overpic}[width=0.5\columnwidth,trim=260 40 260 50,clip]{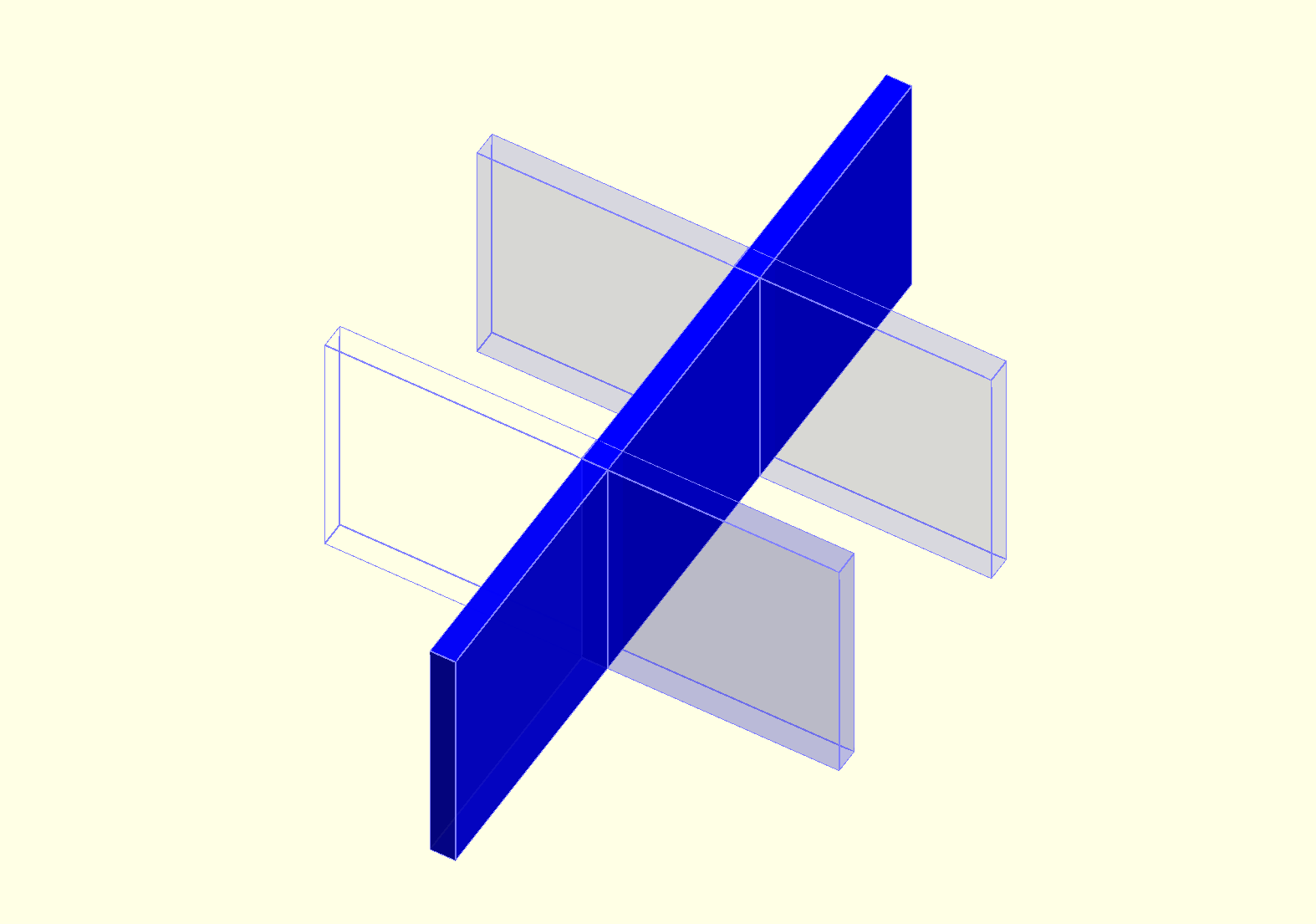}
 \end{overpic}
 \caption{An example of a neighborhood of three-dimensional inputs, for which walls represent the boundariness. The more transparent a wall, the more similar is the neighbors and the lower the boundariness. In this neighborhood, the boundary goes along one axis, with the other axes not showing significant dissimilarity.}
    \label{fig:neighborhood}
\end{figure}

Figure \ref{fig:date2d} shows how a boundary can be extracted for visual analysis in a, here, two-dimensional space. For a fixed non-leap year, we visualize all diversity-measurements nearby those of very high diversity with input parameters \textit{day} and \textit{month}. The color intensity/opaqueness of lines signifies the magnitude of the diversity. Even without access to an oracle, we can see the connected boundary. These visuals can be created on-demand and help the tester to understand areas of input-space with relevant properties. This can further be extended to the visualization in 3d, which is conceptually explained in support of Figure \ref{fig:neighborhood}. There, opacity in a wall highlights the \textit{boundariness} between neighboring inputs/output pairs, suggesting a straight boundary along one axis. The overall space is usually exhaustively large, but with the help of BVE and exploratory techniques that highlight interesting candidate pairs or candidate pair neighborhoods, sub-spaces can be entered and explored visually by free navigation with the calculation of diversity being done on-demand.

\section{Exploring the Boundary}
\label{method}
Even with the visualization of neighborhoods and regions in input search space, and the support of information-theoretical diversity measures, we still do not know where to start to search for potentially wrong boundaries. The search space is potentially huge, with most of it being irrelevant for finding boundaries. Therefore, an exploration requires initial information about potentially interesting areas, which in Julia, \texttt{typemin} and \texttt{typemax}, as introduced above, can provide. They can be validated for the correctness and may serve as starting points for investigation. In other contexts, boundary information can come from developers or testers. We call this entry point an \textit{entrance} to the search. An entrance is a pair of nearby input values that land on the opposite sides of a boundary, e.g., for \texttt{Date}, the pairs $($\texttt{typemin}-$1,$\texttt{typemin}$)$, and $($\texttt{typemax}$,$\texttt{typemax}+$1)$ are examples of entrances.

For data types that contain a \texttt{next} function $\nu$, we can detect boundary candidates by iteratively leveraging $\nu$ with a diversity measure to compare the consecutive values' differences. Outliers are then recommended as boundary candidates for further investigation. In Julia \texttt{Date}, $\nu$ would be the next day function, which in Julia can be obtained adding \texttt{Dates.Day(1)} as in
$$\nu_{Date}(d) = d + \texttt{Dates.Day(1)}$$

We can create a simple algorithm to automatically detect the outliers by iterating through pairs $p=[p_1,\nu(p_1)]$ given an entrance by the user. The general description of that schema is presented in Algorithm \ref{alg:next}. Its mechanics can be illustrated, revisiting Figure \ref{fig:extremes}. Starting with the entrance \texttt{typemax} for $\nu$ provided by the Julia language, the algorithm processes forward until it detects a potential boundary candidate and terminates. As a stop criterion, here, an outlier was declared as being three standard deviations larger than the mean, which is an outlier criterion commonly applied in statistics.\footnote{Other means to do outlier detection are out of the scope of this paper.}

This search is effective, however, not efficient, as it potentially requires passing through the entire search space. It can be used to direct BVE into interesting areas, e.g., for visual exploration. This is exemplified below. Algorithm \ref{alg:next} can be adjusted to detect the lower boundary, which can, for Julia \texttt{Date}, be reached through the \texttt{previous} function $\phi_{Date}(d) = d - \texttt{Dates.Day(1)}$. The BD-algorithm terminates for the visible outlier pairs $b_\phi$ and $b_\nu$ both extremes, respectively.

\begin{algorithm}
{\footnotesize{
\caption{Boundary detection algorithm.}
\label{alg:next}
\begin{algorithmic}
    \REQUIRE next function $\nu$, distance metric $d$, boundary entrance $e = [e_1, e_2]$
    \ENSURE boundary candidate $b^\nu$
    \STATE Pairs = $\emptyset $
    \STATE $p = e$
    \WHILE{$d(p_1, p_2)$ not outlier in $\{d(p'_1, p'_2) | \forall p' \in Pairs \}$}
        \STATE $Pairs = Pairs \bigcup \{p\}$
        \STATE $p = [p_2, \nu(p_2)]$
    \ENDWHILE
    \RETURN $p$
\end{algorithmic}}}
\end{algorithm}



We can further manually validate the correctness of the boundary candidates detected by the algorithm, both visually and through trying them out in code. The from the BD-algorithm returned pairs $b^\phi$ and $b^\nu$ can be found in Table \ref{tab:date}, for $\phi$ as the pair of rows $[8,9]$, and $\nu$ as pair $[4,5]$. The difference taken up by the diversity measure becomes apparent when directly comparing the neighboring outputs in the table. The results for $b^\phi_2$ and $b^\nu_1$ are valid, whereas for $b^\phi_1$ and $b^\nu_2$ they are invalid.

\begin{figure}
\centering
 \begin{overpic}[width=0.7\columnwidth]{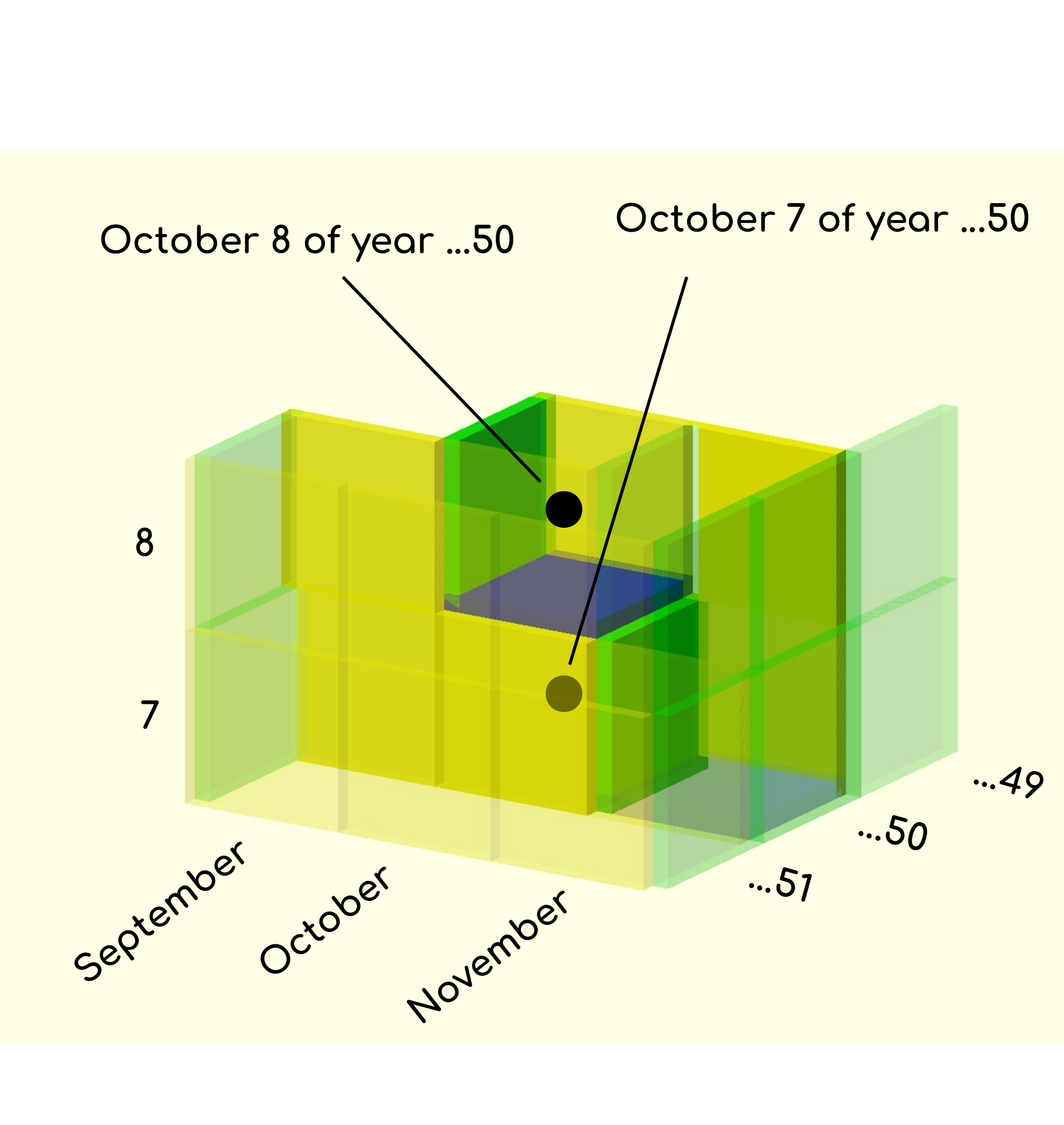}
 \end{overpic}
 \caption{For the exploratory purpose, we look into the boundary space surrounding the pair for which the BD-algorithm terminates given \texttt{typemax}. We see the boundary as the blue plateau in the center with solid coloring, depicting substantial diversity in the input pair it separates. Transparent walls signal that the neighbors are very similar to one another in terms of output diversity.}
    \label{fig:zoom_in}
\end{figure}

The BD-algorithm is single-dimensional, and thus not that helpful for further exploration. However, leveraging the derivative information above, we can graphically investigate the neighborhood of the detected boundary pairs and zoom into or out of interesting regions. Figure \ref{fig:zoom_in} shows the near neighborhood of $b_\nu$; this is where a manual visual investigation may start. The blue plateau in the center depicts the detected boundary pair between October 7 and October 8. Since October 8 has a strongly diverse output that year, an opaque boundary pair is formed with September 8, which can be seen as a solid green wall. The same applies to the pair that splits October 7 on the year ...50 and ...51 (intense yellow). The transparent walls along the different axes are explained by neighbors of high similarity (low boundariness). We expect boundariness to be very low within the valid ranges, and space, therefore, primarily be hollow.

\begin{figure}
\centering
 \begin{overpic}[width=0.8\columnwidth]{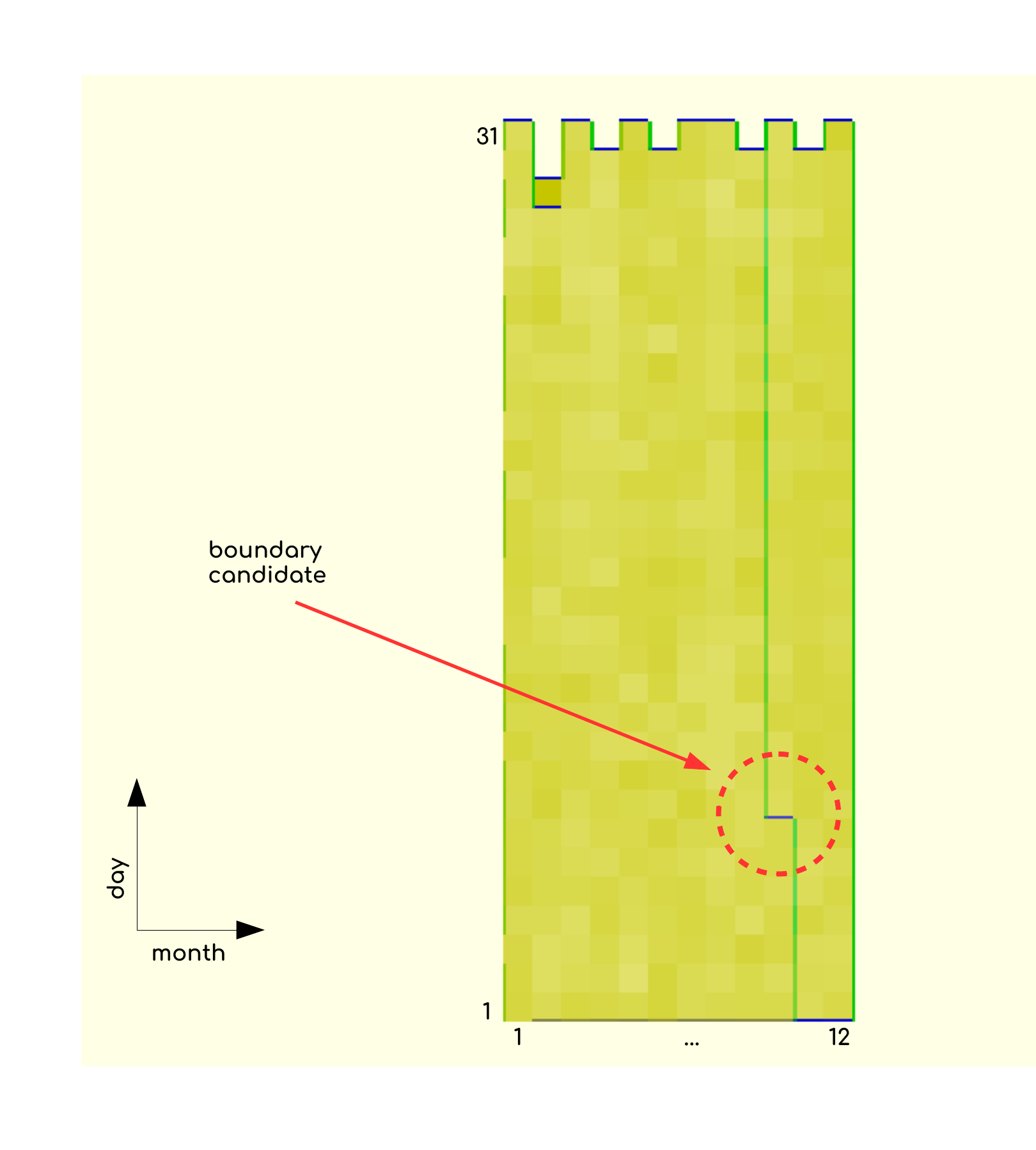}
 \end{overpic}
 \caption{The boundary front stretching beyond the \texttt{typemax} boundary for Julia \texttt{Date} in 2d. The boundary candidate highlighted is the outlier detected by the BD-algorithm.}
    \label{fig:bouundary_candidate_date2d}
\end{figure}

The hollow space can be looked at from different angles. Figure \ref{fig:bouundary_candidate_date2d} zooms out and gives a different perspective, looking at day and month primarily in a seemingly two-dimensional view. The boundary candidate with the plateau $b^\nu$ from Figure \ref{fig:zoom_in} is highlighted for orientation purposes. The outer boundary from Figure \ref{fig:date2d} can be re-identified here, whereas the line crossing September\slash October with a bump in $b_\nu$ may raise curiosity.

\begin{figure}
\centering
 \begin{overpic}[width=0.7\columnwidth]{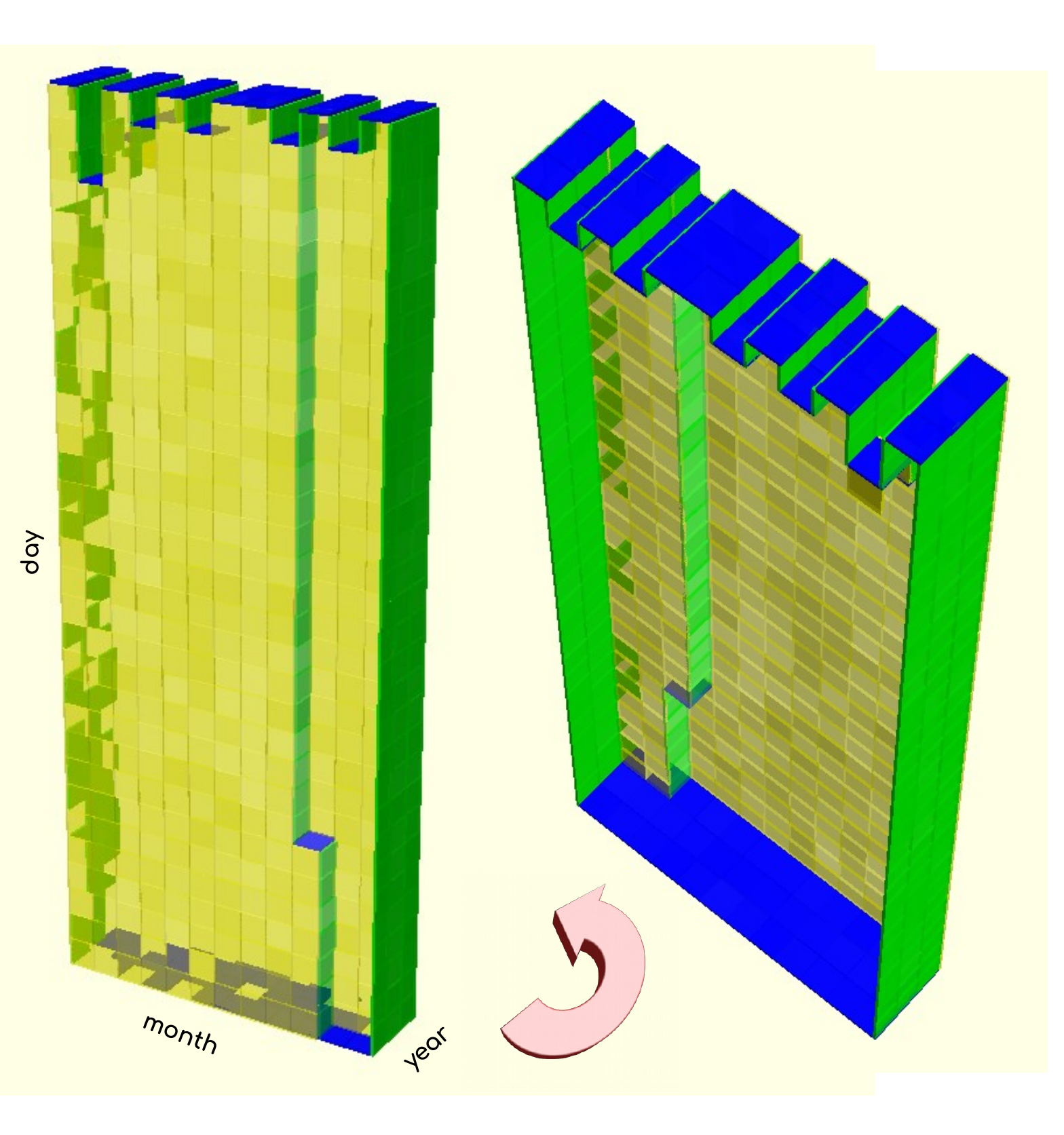}
 \end{overpic}
 \caption{Zooming out from the actual boundary as of row 3 in Table \ref{tab:date}, we see a connected boundary in the sub-space of the inputs (left). Since the valid space extends to October, we see a boundary that splits the October among the valid and invalid spaces. For illustration purposes, the range is limited to four years. The leap year results in a visible deviation from the other years. Rotating the space then (right) shows the hollowness behind the boundary, which represents the similarity in the valid space, and extends far into the valid region, all the way to the other end, which closes not far beyond \texttt{typemin}.}
    \label{fig:zoom_out}
\end{figure}

In Figure \ref{fig:zoom_out} the view is zoomed out to investigate the larger space around the boundaries considering all three dimensions. An investigation in 3d can give clues that search may not capture otherwise. The plateau with the boundary can still be seen in the lower right corner (Figure \ref{fig:zoom_out} left). We further examine the whole shape of the date API in that region, and how almost the entire year after \texttt{typemax} (row in front) is filled with similar days outputs (hollow space) that signal validity up until October 7. The rotation (Figure \ref{fig:zoom_out} right) shows the ample hollow space behind the boundary that leads up to the boundary, signaling the neighboring input/output pairs' relative similarity. This space extends far beyond the visible region, up until the boundary in the vicinity of \texttt{typemax}.

With all the information gathered, detected through the BD-algorithm, and validated visually, in summary, the applied BVE techniques suggest that better candidates for \texttt{typemin} and \texttt{typemax} of the \texttt{Date} type in Julia are: \texttt{typemin(Date)}=\texttt{-252522163911151-07-24} and \texttt{typemax(Date)}=\texttt{252522163911150-10-07}. The current implementation deviates from \texttt{typemin\slash typemax} and further accepts values beyond these boundaries, considered bad design or a bug.

\section{Discussion}
\label{discussion}

In this paper, we argued that the current adoption of boundary value analysis (BVA) and testing (BVT) is limited and then proposed concrete techniques for boundary value exploration (BVE). By building on general ideas for quantifying the distance and diversity between program inputs and outputs, and detecting areas of abrupt changes, we then proposed automated techniques to detect, visualize and, thus, explore boundaries of a Date handling library. Taken together, the application of these techniques allowed us to identify previously unknown inputs to the library where its behavior is questionable. 

We have yet to confirm with the library developers if these constitute expected or only actual boundaries and thus if they indicate bugs or that the specification needs to be updated. However, since the tested library's behavior for these boundaries has recently changed (we performed our tests on Julia version 1.1.1 and then saw a different response on Julia 1.1), there is an indication that the identified inputs were essential to consider.

The crucial future work is to assess the proposed techniques' value in more cases and empirical studies with developers and testers. In particular, such evaluation should include cases where the input and output spaces are complex and structured data types such as XML documents, trees, graphs, since, for them, it is less clear how to visualize changes. For example, if a boundary of implementing a graph-traversing algorithm is heavily tied to the graph's specific structure, it might not be trivial to map differences between the graphs to values (for plotting) or dimensions (for more complex visualizations). While functions that map complex data structures to numbers, e.g., the depth or number of nodes of a tree, might help visualize and then identify boundaries, more non-linear mapping approaches might be needed. Future work should investigate if, for example, methods for dimensionality-reduction can help~\cite{seifert2010stress}.

Further, how to automate BVE, particularly how \textit{entrances} for the search can automatically be identified, shall be investigated in future work. This will include the quest for efficient exploration strategies that scale --- extending our methods will also investigate how to select distance functions given the data types, specification, and\ or implementation. Even if domain knowledge is likely to be critical in this, future work should explore if some general rules can be found or if libraries of (data-type specific) distance functions can be useful and thus reduce the burden on developers and testers to express more specific ones.



\section{Conclusions}
\label{conclusions}
We proposed Boundary Value Exploration as a general concept to help identify boundary values for analysis and testing of software. By utilizing general distance functions, we could detect candidate boundaries and then visualize interesting areas around them, in the case of Date handling library. Overall, our results point to a more automated, agile, and interactive way of analyzing and testing the boundary behavior of software that can lead to both increased effectiveness and efficiency.

{
\tiny
\bibliographystyle{IEEEtran}
\bibliography{main}
}
\end{document}